\documentclass[12pt]{article}
\usepackage{amsfonts,amssymb,amsmath,latexsym}
\newcommand{\be}{\begin{equation}}
\newcommand{\ee}{\end{equation}}
\newcommand{\bea}{\begin{eqnarray}}
\newcommand{\eea}{\end{eqnarray}}
\newcommand{\ba}{\begin{array}}
\newcommand{\ea}{\end{array}}
\newcommand{\bt}{\begin{tabular}}
\newcommand{\et}{\end{tabular}}

\newcommand{\sump}{\mathop{{\sum}'}}
\newcommand{\ti}{\tilde}

\newcommand{\fr}{\frac}
\newcommand{\ci}{\cite}
\newcommand{\cl}{\centerline}
\newcommand{\bs}{\bigskip}
\newcommand{\sms}{\smallskip}
\newcommand{\vs}{\vspace}

\newcommand{\en}{\eqno}

\newcommand{\bbib}{\begin{thebibliography}}
\newcommand{\ebib}{\end{thebibliography}}

\newcommand{\mbb}{\mathbb}

\newcommand{\bm}{\boldmath}
\newcommand{\mb}{\mbox}
\newcommand{\und}{\underline}

\newcommand{\scrs}{\scriptsize}
\begin{document}
\titlepage

\begin{flushright}
\und {\large \it To the 65 anniversary of Vadim Berezinskii}
\end{flushright}
\vs{2cm}

\centerline{\bf CARTAN TORI AND ADE CLASSIFICATION }
\smallskip
\centerline{\bf OF TWO-DIMENSIONAL TOPOLOGICAL}
\sms
\cl{\bf PHASE TRANSITIONS}
\vspace{0.5cm}
\cl{\bf S.A.Bulgadaev}
\vspace{0.5cm}
\centerline{L.D.Landau Institute for Theoretical Physics,}
\cl{Moscow,  RUSSIA}
\centerline{Max Plank Institute for Physics of Complex Systems}
\cl{Dresden, GERMANY}
\vs{0.5cm}
\centerline{A talk presented at the International Conference ICMP-2000}
\cl{17 - 22 July 2000, Imperial College, London, England}

\newpage

\cl{\bf I. INTRODUCTION}
\bs

The topological phase transition (TPT)
or the Berezinskii-Kosterlitz-Thouless (BKT)
phase transition (PT) takes place in two-dimensional
systems with  order parameter
$\psi = e^{2\pi i \varphi} \in {\cal M}=S^1.$
Among them are $XY$-model, superconductors, bose-liquids and many other
systems  \ci{2,18,14,12,21}.
$XY$-model on a lattice is defined as follows
$$
{\cal Z}_{XY} = \sum \exp (-\beta {\cal H}), \quad
$$
$$
{\cal H}= -  \fr{1}{2}\sum_{<i,j>} J (\psi_i \psi_j^* + c.c.)=
- J \sum_{<i,j>} \cos(\varphi_i - \varphi_j).
\en(1)
$$
Its continuous variant is a nonlinear $\sigma$-model (NSM) on $S^1$
$$
{\cal Z}_{NS}= \int D\varphi e^{-{\cal S}[\varphi]}, \quad
{\cal S}_{NS}= \fr{1}{2\alpha}\int |\partial \psi|^2 d^2 x =
\fr{1}{2\alpha} \int (\partial \varphi)^2 d^2 x,
\en(2)
$$
$$
\alpha \simeq T/2J.
$$
A circle $S^1$ has a nontrivial  homotopic group $\pi_1$
$$\pi_1(S^1)=\mbb {Z}.
\en(3)
$$
Due to this fact the topologically stable excitations, vortices,
are possible in these systems.
One vortex solution has a form (at large distancies)
$$
\varphi({\bf x})= \fr{1}{\pi} \arctan \fr{y}{x},
\quad {\bf x} =(x,y) \in \mbb {R}^2,
\en(4)
$$
An account of vortices means that theory must be considered on the
covering space $\mbb {R}$ of the circle $S^1 = \mbb {R}/\mbb {Z}.$
The energy of one "vortex" 
is logarithmically divergent, but
the energy $E_N$ of $N$ vortices with the full topological
charge $e = \sum _{i=1}^N e_i = 0$ is finite and equals
$$
E_N= \fr{2\pi}{2\alpha}\sum_{i\ne k}^N e_i e_k
\ln \fr{|{\bf x}_i-{\bf x}_k|}{a} + C(a) \sum_i^N e_i^2,
\en(5)
$$
here $C(a)$ is some nonuniversal constant, determining "self-energy"
(or core energy)
of vortices and depending on type of core regularization.

The partition function ${\cal Z}_{XY}$ can be
approximated by product of two partition functions \ci{14,12}
$$
{\cal Z}_{XY}\simeq {\cal Z}_{sw} {\cal Z}_{CG},
\en(6)
$$
where ${\cal Z}_{CG}$  is the grand partition function
of dilute Coulomb gase (CG) of topological exitations with minimal
charges $e=\pm1$ and
${\cal Z}_{sw}$ is the partition function of
{\emph{free}} "spin-waves".
${\cal Z}_{CG}$  can be represented, in its turn, in the form
of effective  field theory with  sine-Gordon
 (SG) action ${\cal S}_{SG}$ \ci{12,21}
$$
{\cal Z}_{CG}= {\cal Z}_{SG} = \int D\phi e^{-{\cal S}_{SG}[\phi]} ,
\en(7)
$$
$$
{\cal S}_{SG}= \int \left[ \fr{1}{2\alpha}(\partial \phi)^2
- 2 \mu^2 \cos \phi\right] d^2 x
$$
The TPT takes place in the system of  vortices and can be described by
the effective SG theory.
This system has two different phases:

\bs

1) \underline {\bf high-T phase}: plasma-like, massive, with finite
correlation length with essential singularity at $T_c$ \ci{14}
$$
\xi \sim a\exp (A\tau^{-\nu}), \quad \nu=\fr{1}{2},
\en(8)
$$
$$
\tau = \fr{T-T_c}{T_c} \to 0, \quad T_c \approx \pi J;
$$

2) \underline {\bf low-T phase}: dielectric, massless, with infinite
correlation length and algebraically falling correlations \ci{20,2,18}.

\sms

\und {\bf Question}:

\sms

Are there any possible generalizations on systems with more
complicate group $\pi_1$
and other types of critical behaviour?

\bs

The answer is nontrivial. The simplest generalization of circle $S^1$
is a torus $T^n$ with the homotopy group
$$
\pi_1(T^n) = \bigoplus^n_{i = 1}{\mbb Z}_i = {\mbb Z}^n,
\en(9)
$$
where $i$-th component describes maps of the boundary $S^1$ into
$i$-th circle of $T^n.$

The maps into different components cannot
be transformed into or annihilate each other.
Consequently, one can introduce in $\pi_1(T^n)$ and in space of
corresponding topological charges a vector structure:
a vector basis and a metric.
In case of $T^n$ it is an usual euclidean structure with canonical
basis $\{{\bf e}_i\}$ and metrics
$g_{ik} = \sum_{a=1}^n e^a_ie^a_k = \delta_{ik}.$ Then the topological charges,
corresponding to different $S^1_i,$
do not interact!
Thus, the theories with ${\cal M} = T^n$ simply replicate
the case ${\cal M} = S^1$  and reduce to it.

Moreover, all tori $T_L = {\mbb R}^n/\mbb {L}$,
$$
\mbb {L} = \sum_{i=1}^n n_i {\bf e}_i, \,\, n_i \in {\mbb Z}_i, \,\,
{\bf e}_i \in \{{\bf e}_i\}_L, \quad g_{ik}=\sum_{a=1}^n e^a_i e^a_k,
\en(10)
$$
where $\mbb {L}$ is $n$-dimensional lattice in ${\mbb R}^n$
($\{{\bf e}_i\}_L, \,i=1,...,n$ forms a basis of lattice $\mbb {L}$),
and
$g_{ik}$ is an effective metric determined by the theory action,
reduce in the same sence to the case $S^1$ also \ci{4}.
It means that

\sms

\und {the torus properties have some hardness (or stability):}

\sms

\und {the smooth deformations in general position do not change them.}

\sms

This fact is connected with corresponding deformation of effective
metric in space of topological charges.
In terms of critical properties of TPT it means that for all
systems, having such tori as a vacuum space, the TPT will have the
same critical properties \ci{4}.

\bs

{\bf Proposal:}

\bs

\framebox{\parbox[1.5\height]{13.7cm}{
For obtaining nontrivial generalizations one must consider
NSM on the degenerated tori, in particular, on the Cartan tori $T_G$
of the simple compact Lie groups $G.$
}}

\bs

\cl{\bf II. CARTAN SUBGROUP AND DEGENERATED  TORI}

\bs

The Cartan torus $T_G,$ the maximal abelian
subgroup of group $G,$ consists of elements
$$
{\bf g} = e^{2\pi i({\bf H} \mb {\scrs \bm $\phi$})},\;
{\bf H} = \{H_1,...,H_n\} \in {\cal C}, \; [H_i,H_j] = 0 ,
$$
where $n$ is a rank of $G,$
${\cal C}$ is a maximal commutative Cartan
subalgebra of the Lie algebra ${\cal G}$ of the group $G.$
It is assumed here that
$({\bf H} \mb {\bm $\phi$})$ is an usual euclidean scalar product.
All $H_i$ can be diagonalized simultaneously.
Their eigenvalues,  the weights  ${\bf w}$ (or quantum numbers),
depend on the concrete
representation of $G$ and ${\cal C}.$

All possible weights  ${\bf w}$ of the simply connected
group $G$ (or of the universal covering group $\ti G$ of
the non-simply connected group $G$)
form a lattice of weights $L_{w}$.
In this basis  all $H_i$ (and any element ${\bf g}\in T_G$) get
a diagonal form
$$
{\bf g}_{\tau}=
diag(e^{2\pi i({\bf w}_1 \mb {\scrs \bm $\phi$})},...,
e^{2\pi i({\bf w}_p \mb {\scrs \bm $\phi$})})
\en(11)
$$

The main differences of this form from the usual representation of
$T_L$ type tori are:

1) a dimension of diagonal matrices coincides with dimension
$p$ of $\tau$-representation, which is usually larger, than rank of $G$;

2) the set of weights
$\{{\bf w}\}_{\tau}$ has a discrete Weyl (or crystallographic) symmetry,
which results in the next two
properties
$$
\sum_{a=1}^p {\bf w}_a = 0, \quad
g_{ik} = \sum _{a=1}^p w^a_i w^a_k = B_{\tau}\delta_{ik},
\en(12)
$$
where constant $B_{\tau}$ depends on representation.
Now the effective metrics $g_{ik}$  is
proportional to the euclidean one.

Tori $T_G$ can be considered
as appropriately constrained (reduced)
torus $T^N = T_{U(N)}$ with large enough $N$.
They give the examples of degenerated tori ${\cal T}_L$,
which we define, in general case,
by diagonal matrices of form (11), containing all minimal vectors of lattice
$\mbb {L}.$ A dimension of these matrices $p$ equals to the number
of all minimal vectors of lattice $\mbb {L},$ which is usually
larger than dimension of lattice.

In the form (11) all ${\bf g}\in T_G$
are periodic with a lattice of periods $L_{\tau}^{t}$,
inverse to the lattice $L_{\tau}$, generated
by weights ${\bf w}_a (a=1,...,p)$ of $\tau$-representation.
$L_{\tau}^{t}$ forms a set of all topological charges of
$\tau$-representation of $T_G.$

The  lattice $L^t_{\tau}$ satisfies the next restriction
$$
L_{w^*} \supseteq L^t_{\tau} \supseteq L_v.
$$
where $L_{w^*}$ is a weight lattice  of dual group $G^*,$
$L_v$ is a lattice of dual roots.
For $\tau = min$ a lattice $L^t_{\tau} = L_v,$
for $\tau =ad$ a lattice $L^t_{\tau} = L_{w^*}.$
The lattices $L_v$ and $L_{w^*}$ differ by a factor, which is isomorphic
to the centre $Z_G$ of group $G$
$$
L_{w^*} / L_v = Z_G.
$$
Thus the set of minimal topological charges $\{{\bf q}\}_{\tau}$
can vary from the set of minimal vectors
of the weight lattice till that of the dual root lattice.
All possible cases are determined by
subgroups of the centre $Z_G.$ For groups $G$ with $Z_G=1$ the lattices
$L_v$ and $L_{w^*}$ coincide, i.e. they are self-dual ($G=E_8$).

\bs

\fbox{\parbox[c][1.2\height]{13.7cm}{
When $L^t_{\tau} = L_w$ (or $L^t_{\tau} = L_{w^*}$)
all weights (i.e. quantum numbers) of group $G$ (or $G^*$)
can be reproduced as the vector
topological charges of vortices!}}

\bs

Analogously, a lattice of topological charges $L^t_L$ of degenerated torus
${\cal T}_L$  belongs to a reciprocal lattice $\mbb {L}^{-1}.$ Since
the integer-valued lattices $L$ also belong to their own inverse lattices
$\mbb {L}^{-1},$  their lattices of topological charges, in general,
are even larger then $\mbb L:$
$L^t_L \supseteq  \mbb {L}.$ For this reason they can contain
{\it fractional} (in this basis) topological charges.
Only for degenerated tori ${\cal T}_L,$
connected with self-dual lattices, $L^t_L$ exactly coincide with $L.$
Consequently, for tori associated with the integer-valued lattices
all their "quantum numbers" always have a topological interpretation.

\bs

\cl{\bf III. NS-MODELS ON $T_G$, DUALITY}
\sms

\cl{\bf AND EFFECTIVE THEORIES}

\bs

The euclidean two-dimensional NSM on $T_G$,
have the following form
$$
{\cal S} = \frac{1}{2\alpha} \int d^2x Tr_{\tau}({\bf t}^{-1}_{\nu}
{\bf t}_{\nu}) =
\frac{(2\pi)^2}{2\alpha} \int d^2x Tr_{\tau}({\bf H}
{\mb {\bm $\varphi$}}_{\nu})^2
\en(13)
$$
$$
= \frac{(2\pi)^2}{2\alpha}B_{\tau}\int d^2x ({\mb {\bm $\varphi$}}_{\nu})^2,
$$
where ${\mb {\bm $\varphi$}}_{\nu} = \partial_{\nu} \mb {\bm $\varphi$},
\, \nu = 1,2.$
An including of a factor $B_{\tau}$
into trace $Tr_{\tau}$
gives a canonical euclidean metric in space of topological charges.

These theories are invariant under direct product of right (R) and left (L)
groups $N_G^{R(L)},$ which are a semi-direct product of $T_G$ and $W_G$
$$
N_G = T_G \times W_G.
\en(14)
$$
The group $N_G \in G$ is called a normalizator of $T_G$ and is a symmetry group
of torus $T_G,$ thus the theories (13) can be considered also
as chiral NSM on group $G$ with  symmetry breaking $G \searrow N_G.$
The NSM on $T_G$ have properties analogouos to those of $XY$-model:

1) a zero beta-functions $\beta(\alpha)$ due to flatness of $T_G;$

2) non-trivial homotopy group $\pi_1$ and corresponding vortex
solutions with topological charges ${\bf q} \in L^t_{\tau}$.

In quasi-classical approximation (or in low T expansion)
a partition function of the $\sigma$-model
on $T_G$
can be represented as a grand partition function of classical
neutral Coulomb gas of vortices with  topological charges
${\bf q}_i \in \{{\bf q}\}_{\tau}$
$$
{\cal Z}= {\cal Z}_0 {\cal Z}_{CG},\quad
{\cal Z}_{CG}= \sum_{N=0}^{\infty} \frac{\mu^{2N}}{N!} \sump_{\{{\bf q}\}}
{\cal Z}_N(\{{\bf q}\}|\beta).
\en(15)
$$
Here
${\cal Z}_0$ is a partition function of free massless isovectorial
boson field which corresponds to "spin waves" of $XY$-model.

This gives an embedding of the compact $\sigma$-models on $T_G$ into
noncompact generalized SG theories
$$
{\cal Z}_{CG} = \int D\mb {\bm $\phi$} e^{-{\cal S}_{eff}}, \quad
{\cal S}_{eff} = \int \fr{1}{2\beta}(\partial \mb {\bm $\phi$})^2
-\mu^2 V(\mb {\bm $\phi$}),
\en(16)
$$
$$
V(\mb {\bm $\phi$}) =
\sum_{\{{\bf q}\}} \exp i({\bf q} \mb {\bm $\phi$}).
$$
where $\sum_{\{{\bf q}\}}$ goes over the set of minimal topological
charges, and $\mb {\bm $\phi$} \in {\mbb R}^n$.
The initial NSM correspond to some relation between
parameters $\mu$ and $\beta$.

The account of vortices reduces the initial symmetry group
$N_G$ into discrete dual group
$W_{G^*} \times L_q^{-1}$
($W_{G^*}$ is a dual Weyl group, $L_q^{-1}$ is
a periodicity lattice   of potential $V$).
This dual group generalizes
the dual group $Z_2 \times {\mbb Z}$  of $XY$-model.

Thus, in this semiclassical and long wavelength approximation

\bs

\fbox{
\parbox[c][1.2\height]{13.7cm}{{\it Compact}
theory on a torus $T_G$ with {\it continuous} symmetry $N_G$ appears
equivalent (modulo ${\cal Z}_0$)
to {\it noncompact}
theory with periodic potential and an {\it infinite discrete} symmetry.}}

\bs

In case $\tau = ad$ the generalized SG theories can describe
other systems with symmetry $G$ broken to $N_G$ \ci{3}.

\bs

\cl{\bf IV.  ADE LATTICES, TPT AND SYMMETRIES}

\bs

The critical properties of the BKT type PT can be determined
by renorm-group method \ci{14,21}.
The new critical properties  appear only in case,
when each  vector ${\bf q} \in \{{\bf q}\}$
can be represented as a sum of two other vectors.
This condition concides with a definition
of the root systems $\{{\bf r}\}$ of simple groups from series
$A,D,E$  or of the root sets of the even
integer-valued  lattices of $\mbb {A,D,E}$ types.
Moreover, the sets of minimal roots (and  minimal weights) of all simple
groups belong to four series of the integer-valued (in appropriate
scale) lattices $\mbb {A,D,E,Z}.$
$Z_n$ is an example of the odd self-dual (or unimodular) lattices
and contains a minimal vectors with norm equal 1, while the series
$A_n, D_n, E_n$ belong to the even lattices with minimal norm equal 2.

Each root set is characterised by the Coxeter number $h_G$
$$
h_G=\frac{\mbox{(number of all roots)}}{\mbox{(rank of group)}},
$$
$$
h_{A_n} = n+1,\; h_{D_n} = 2(n-1),\; h_{E_{6,7,8}}= 12,18,30.
$$

\bs

\fbox{
\parbox[c][1.2\height]{13.7cm}{
All coefficients of RG equations are expressed only through
the Coxeter numbers $h_G$ }}

\bs

The phase diagram have two separatrices
with next declinations ($u_1$ corresponds to the phase separation line):
$u_{1,2}= (1/\pi h_G, -1/2\pi).$

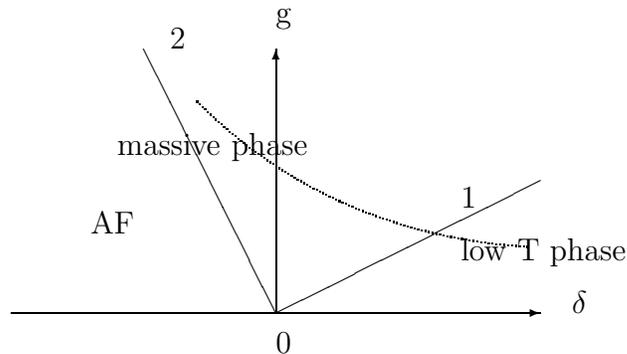
\begin{picture}(400,150)(-75,0)
\put(100,0){\vector(1,0){100}}
\put(100,0){\vector(0,1){100}}
\put(100,0){\line(-1,0){100}}
\put(100,0){\line(2,1){100}}
\put(100,0){\line(-1,2){50}}
\put(170,20){low T phase}
\put(40,60){massive phase}
\put(30,30){AF}
\put(210,0){$\delta$}
\put(100,110){g}
\put(100,-15){0}
\put(60,100){2}
\put(170,40){1}
\qbezier[100](70,80)(120,30)(195,25)
\end{picture}

\bs

The dashed line of the initial values  corresponds to the initial
$\sigma$-model. The critical exponent $\nu_G$
is inverse to the Lyapunov index of the separatrix 1 and equals
$$
\nu_G = 2/(2+h_G).
$$
It can take  the next values \ci{3,4}
\sms
$$
\ba{|c|c|c|c|c|c|c|c|c|c|}
\hline
{} & {} & {} & {} & {} & {} & {} & {} & {} & {}\cr
G & A_n & B_n & C_n & D_n & G_2 & F_4 & E_6 & E_7 & E_8\cr
{} & {} & {} & {} & {} & {} & {} & {} & {} & {}\cr
\hline
{} & {} & {} & {} & {} & {} & {} & {} & {} & {} \cr
\nu_G & \fr{2}{n+3} & \fr{1}{n} & \fr{1}{2} & \fr{1}{n} & \fr{2}{5}
& \fr{1}{4} & \fr{1}{7} & \fr{1}{10} & \fr{1}{16} \cr
{} & {} & {} & {} & {} & {} & {} & {} & {} & {} \cr
\hline
\ea
$$

\bs

The critical properties of some systems with different symmetries
can coincide due to coincidence of their $h_G$.

\bs

\underline {\bf Low-T  phase properties}

\bs

The low-temperature phase is described by effective free field theory
with a renomalized
"temperature" $\bar \beta,$ depending on initial values $\beta_0.$
At the PT point (where  $\bar \beta = \beta^{*} = 8\pi/r^2 = 4\pi$) an
additional logarithmic factors, related with the "null charge" behaviour
of the renormalized parameters on the critical separatrix
(the phase separation line), can appear.


Free-like behaviour of the low-temperature phase
(except logarithmic corrections at criticality)
admits for its description the conformal
field theories with \underline {integer} central charges $C = n$,
instead of PT points of two-dimensional systems with discrete symmetries,
described by conformal theories with
\underline {rational} central charges \ci{1,8}.
The BKT  PT can be considered as the limiting
case $k\to \infty$ (where $k$ is a level) of the sequence of
minimal conformal theories with $C=1-6/(k+1)(k+2)$ \ci{19}.
Analogously, the TPT in $\sigma$-models on $T_G$ are the limiting cases
of unitary minimal conformal theories, connected with
conformal $W$-algebras \ci{9}.
There exists
a puzzling coincidence of  $\nu_G$ with "screening" factor
in formulae for central charges of the affine Lie algebras
$\hat {\mathfrak{G}}$ \ci{13} at level $k=2$
(though $T_G$ corresponds to $k=1$)
$$
C_k = \frac{k}{k+h_G} dim G
$$
and of the coset realization of the minimal unitary conformal
models \ci{10} at level $k=1$
$$
C_k =
r\left(1-\frac{h_G(h_G +1)}{(k+h_G)(k+h_G +1)}\right).
$$

\bs

\underline {\bf Properties of massive phase}

\bs

In this phase all additional vector charges will be shielded like in plasma.
All excitations are massive.
There is another  enlargement
of the isotopic symmetry of the initial NSM on  separatrix 2.
$\sigma$-model on $T_G$ has at classical level two continuous symmetries:
1) scale (or conformal) symmetry, 2) isotopic global symmetry
$N_G = T_G \times W_G.$
Both symmetries are spontaneously broken in IR region
by vortices. For this reason $\sigma$-model has in massive phase
a finite correlation length $\xi \sim  m^{-1}$,
where  $m$ is some characteristic mass scale of the theory.
This mass can depend on the coupling constant
$\beta.$

On separatrix 2 one obtaines
$$
m \sim\ \Lambda \exp{(-1/2\pi gh_G)}, \quad \Lambda \sim  a^{-1}.
$$
This mass scale is defined only by $K_G \sim h_G$ (note that here $G=ADE$)
and coincides in main approximation with those for chiral models on
groups $G$ \ci{17,16,7} and for fermionic models with symmetry group
$G$ \ci{7}.

\bs

\fbox{\parbox[c][1.2\height]{13.7cm}{
Mass scale of NSM on $T_G$ coincides  on separatrix 2
with that for all $G$-invariant theories,
the chiral as well as fermionic (at least for $G=A,D,E$).
}}

\bs

It means a possible restoration of the full isotopic symmetry group $G$
in $\sigma$-models with symmetry group $N_G$ in massive phase.

\bs

\und{\bf ADE classification}
\bs

There are a number of other integer-valued  lattices, which can serve
as a lattice of topological charges.
Their classification is not completed at present (except some
low-dimensional cases) \ci{6}.
But, if one confines  himself with NSM on tori ${\cal T}_L$
with integer-valued lattices of
topological charges (their importance was noted above),
then all possible types of critical behaviour will belong only
to $ADEZ$ series.  This conclusion follows from
the Witt theorem, proving that
the minimal vector sets of any integer-valued lattice must be a
direct sum of the root systems \ci{6}.
But the last can be only of $A,D,E,Z$ types.
Consequently, all NS-models on tori ${\cal T}_L$  with integer-valued
lattices of topological charges $L^t_L$ with minimal norm equal 1
can have critical properties only of
$XY$-model (or of $Z^n$ lattice) type, while all
NS-models on tori ${\cal T}_L$  with integer-valued lattices $L^t_L$
with minimal norm equal 2 can have critical properties only of $A,D,E$
lattice types. In this relation it is worth to note that analogous
ADE classifications take place in the  theory of singularities \ci{}
and in the string theory \ci{}.
.
In general case of integer-valued lattices the different components
of the minimal vector sets
can have different $h_G.$ Then one can have in NSM
on tori ${\cal T}_L$  with general integer-valued lattice $L^t_L$
a series of PTP,
taking place separately in each component (with critical properties,
depending on $h_G$).
For even self-dual lattices with minimal norm 2
all components must have the same Coxeter number \ci{6}, and,
consequently, the PTP in all components take place simultaneously \ci{5}.

The TPT in NSM on degenerated tori can have an application to
the description of partial space decompactification in string theories \ci{5}.

\bs

\cl{\bf V. CONCLUSIONS}

\bs

1. It is shown that one must consider deformed tori for obtaining
the interacting vector topological charges.

2. Vector topological charges form a lattice and in some cases can
reproduce all quantum numbers of the corresponding groups.

3. A sequence of approximately equivalent
transformations of 2d  models is constructed

\bs

\fbox{\parbox[c][2\height]{13.7cm}{\hspace{2cm} General GL Theory $\to$
NS Model $\to$
TopExcGas

\cl{$\to$ General SG Theory}}}
\bs

It simplifies a problem and extracts all necessary long-wave properies
of these theories! Here last theory has a pure group-theoretical
structure and is universal for  whole class of theories with the same
symmetries.

4. All critical properties are classified by integer-valued lattices
from series $\mbb {A,D,E,Z}$ and are characterised by the corresponding
Coxeter numbers.

5. The possible scenarios of the dynamical enlargement of the initial
internal symmetry groups is discussed.

6. Some  applications of these TPT for cosmological and string
theories is proposed.
\bs

I would like to thank the organisers of the conference for
the opportunity to present this talk and support.

This work was supported by RFBR (grant N 00-15-96579) and
Dresdener Max Plank Institute for Physics of Complex Systems.

 \end{document}